% last revised Aug 3, 1997
%
\let\includefigures=\iffalse
%
% the following is to use blackboard bold fonts --
\let\useblackboard=\iftrue
%
% activate this if you don't have them.
%\let\useblackboard=\iffalse
%
% You might also need to remove this line.
\newfam\black
\input harvmac
\includefigures
\message{If you do not have epsf.tex (to include figures),}
\message{change the option at the top of the tex file.}
\input epsf
\def\figin{\epsfcheck\figin}\def\figins{\epsfcheck\figins}
\def\epsfcheck{\ifx\epsfbox\UnDeFiNeD
\message{(NO epsf.tex, FIGURES WILL BE IGNORED)}
\gdef\figin##1{\vskip2in}\gdef\figins##1{\hskip.5in}% blank space instead
\else\message{(FIGURES WILL BE INCLUDED)}%
\gdef\figin##1{##1}\gdef\figins##1{##1}\fi}
\def\DefWarn#1{}
\def\figinsert{\goodbreak\midinsert}
\def\ifig#1#2#3{\DefWarn#1\xdef#1{fig.~\the\figno}
\writedef{#1\leftbracket fig.\noexpand~\the\figno}%
\figinsert\figin{\centerline{#3}}\medskip\centerline{\vbox{\baselineskip12pt
\advance\hsize by -1truein\noindent\footnotefont{\bf Fig.~\the\figno:} #2}}
\bigskip\endinsert\global\advance\figno by1}
%%%
\else
\def\ifig#1#2#3{\xdef#1{fig.~\the\figno}
\writedef{#1\leftbracket fig.\noexpand~\the\figno}%
%\figinsert\figin{\centerline{#3}}\medskip\centerline{\vbox{\baselineskip12pt
%\advance\hsize by -1truein\noindent\footnotefont{\bf Fig.~\the\figno:} #2}}
%\bigskip\endinsert
\global\advance\figno by1}
\fi
\useblackboard
\message{If you do not have msbm (blackboard bold) fonts,}
\message{change the option at the top of the tex file.}
\font\blackboard=msbm10 scaled \magstep1
\font\blackboards=msbm7
\font\blackboardss=msbm5
\textfont\black=\blackboard
\scriptfont\black=\blackboards
\scriptscriptfont\black=\blackboardss

\else

\fi
% *************************************
%\draft
%
\def\yboxit#1#2{\vbox{\hrule height #1 \hbox{\vrule width #1
\vbox{#2}\vrule width #1 }\hrule height #1 }}
\def\fillbox#1{\hbox to #1{\vbox to #1{\vfil}\hfil}}
\def\ybox{{\lower 1.3pt \yboxit{0.4pt}{\fillbox{8pt}}\hskip-0.2pt}}

\def\comments#1{}
\def\cc{{\rm c.c.}}

\def\p{\partial}

\def\eps{\epsilon}
\def\half{{1\over 2}}
\def\Tr{{{\rm Tr~ }}}
\def\STr{{{\rm STr~ }}}
\def\Str{{{\rm STr~ }}}
\def\tr{{\rm Tr\ }}
\def\det{{\rm det\ }}
\def\Re{{\rm Re\hskip0.1em}}

\def\CM{{\cal M}}
\def\CN{{\cal N}}

\def\CL{{\cal L}}

\def\CW{{\cal W}}
\def\CX{{\cal X}}

\def\nl{\hfill\break}

\def\ap{\alpha'}

\def\II{\relax{I\kern-.10em I}}

\def\IZ{\relax\ifmmode\mathchoice
{\hbox{\cmss Z\kern-.4em Z}}{\hbox{\cmss Z\kern-.4em Z}}
{\lower.9pt\hbox{\cmsss Z\kern-.4em Z}}
{\lower1.2pt\hbox{\cmsss Z\kern-.4em Z}}\else{\cmss Z\kern-.4em
Z}\fi}
\def\IB{\relax{\rm I\kern-.18em B}}
\def\IC{{\relax\hbox{$\inbar\kern-.3em{\rm C}$}}}
\def\ID{\relax{\rm I\kern-.18em D}}
\def\IE{\relax{\rm I\kern-.18em E}}
\def\IF{\relax{\rm I\kern-.18em F}}
\def\IG{\relax\hbox{$\inbar\kern-.3em{\rm G}$}}
\def\IGa{\relax\hbox{${\rm I}\kern-.18em\Gamma$}}
\def\IH{\relax{\rm I\kern-.18em H}}
\def\II{\relax{\rm I\kern-.18em I}}
\def\IK{\relax{\rm I\kern-.18em K}}
\def\IP{\relax{\rm I\kern-.18em P}}
%\def\IX{\relax{\rm X\kern-.01em X}}
%this doesn't work

%

\def\inbar{\,\vrule height1.5ex width.4pt depth0pt}

\def\p{\partial}

\font\cmss=cmss10 \font\cmsss=cmss10 at 7pt
\def\IR{\relax{\rm I\kern-.18em R}}

\def\bi{{\bar{i}}}
\def\bj{{\bar{j}}}
\def\bk{{\bar{k}}}
\def\bl{{\bar{l}}}
\def\bm{{\bar{m}}}
\def\bn{{\bar{n}}}
\def\bq {{\bar{q}}}
\def\bx {{\bar{x}}}
\def\by {{\bar{y}}}
\def\bz {{\bar{z}}}
\def\bZ{{\bar Z}}

\def\BR{\IR}

\def\BR{\IR}
\def\BC{\IC}

\def\lp10{l_P^{10}}
\def\lp11{l_P^{11}}
\def\R11{R_{11}}

\Title{\vbox{\baselineskip12pt\hbox{hep-th/9708012}
\hbox{LBNL-40515}
\hbox{RU-97-67}
\hbox{UCB-PTH-97/36}
\hbox{UTMS 97-??}}}
{\vbox{
\centerline{D-brane Actions on K\"ahler Manifolds} }}
\centerline{Michael R. Douglas$^1$, Akishi Kato$^{2,3}$  and Hirosi
Ooguri$^{2,4}$}
\medskip
\centerline{$^1$Department of Physics and Astronomy, Rutgers University }
\centerline{Piscataway, NJ 08855--0849, U.S.A.}
\centerline{\tt mrd@physics.rutgers.edu}
\medskip
\centerline{$^2$Theoretical Physics Group, Mail Stop 50A-5101}
\centerline{Lawrence Berkeley National Laboratory, Berkeley, CA 94720,
U.S.A.}
\centerline{\tt akishi@ms.u-tokyo.ac.jp, ooguri@mack.lbl.gov}
\medskip
\centerline{$^3$Department of Mathematical Sciences, University of Tokyo}
\centerline{Komaba 3-8-1, Meguro-ku, Tokyo 153, Japan}
\medskip
\centerline{$^4$Department of Physics, University of California at
Berkeley}
\centerline{366 Le Conte Hall, Berkeley, CA 94720-7300, U.S.A.}
\bigskip
\noindent
We consider actions for $N$ D-branes at points in a general
K\"ahler manifold, which satisfy the axioms of D-geometry,
and could be used as starting points for defining Matrix theory
in curved space.

We show that the axioms cannot be satisfied unless the metric
is Ricci flat, and argue that such actions do exist when the metric
is Ricci flat.  This may provide an argument for Ricci flatness in
Matrix theory.

\Date{August 1997}
%\draft
%
\lref\dlp{J.~Dai, R.~G.~Leigh and J.~Polchinski,
``New connections between string theories,'' Mod. Phys. Lett. {\bf A4}
(1989) 2073;
J.~Polchinski, ``Dirichlet branes and Ramond-Ramond charges,''
Phys.~Rev.~Lett.~{\bf 75} (1995) 4724-4727,
hep-th/9510017.}
\lref\leigh{R. Leigh, ``Dirac-Born-Infeld action from Dirichlet sigma
model,'' Mod. Phys. Lett. {\bf A4} (1989) 2767.}
\lref\witten{E. Witten, Nucl. Phys. B443 (1995) 85, hep-th/9503124.}
\lref\dos{M. R. Douglas, H. Ooguri and S. H. Shenker,
``Issues in M(atrix) theory compactification,''
Phys. Lett. {\bf B402} (1997) 36-42, hep-th/9702203.}
\lref\dg{M. R. Douglas and B. R. Greene, ``Metrics on D-brane Orbifolds,''
hep-th/9707214.}
\lref\dstrings{M. R. Douglas, ``D-branes and Matrix theory in Curved Space,''
lecture at Strings '97, hep-th/9707228.}
\lref\dcurve{M. R. Douglas, ``D-branes in Curved Space,''
hep-th/9702048.}
\lref\polgeo{J. Polchinski, hep-th/9611050.}
\lref\egs{M. R. Douglas, hep-th/9612126.}
\lref\bfss{T. Banks, W. Fischler, S. H. Shenker and L. Susskind,
``M Theory As A Matrix Model: A Conjecture,'' Phys.Rev. {\bf D55} (1997)
5112, hep-th/9610043.}
\lref\banks{T. Banks, ``The State of Matrix Theory,'' hep-th/9706168.}
\lref\DKPS{M. R. Douglas, D. Kabat, P. Pouliot and S. Shenker,
hep-th/9608024.}
\lref\tseytlin{A. Tseytlin, hep-th/9701125.}
\lref\sen{A. Sen, ``BPS States on a Three Brane Probe,'' 
Phys.Rev. {\bf D55} (1997) 2501, hep-th/9608005.}
\lref\connes{A. Connes, {\it Noncommutative Geometry,} Academic Press, 1994.}
\lref\hull{C. Hull, A. Karlhede, U. Lindstr\"om and M. Ro{\v c}ek,
Nucl. Phys. B266 (1986) 1.}
\lref\vafa{C. Vafa, hep-th/9602022.}
\lref\taylor{W. Taylor, hep-th/9611042.}
\lref\dewit{B. de Wit, J. Hoppe and H. Nicolai,
Nucl.Phys. {\bf B 305 [FS 23]} (1988) 545.}
\lref\hadamard{J. Hadamard, {\it Lectures on Cauchy's Problem in Linear
Partial Differential Equations}, Yale University Press, 1923 and
Dover, 1952;
as discussed in J. J. Duistermaat, {\it The Heat Kernel Lefschetz
Fixed Point Formula for the Spin-C Dirac Operator,} Birkh\"auser 1996.}
\lref\rozwit{L. Rozansky and E. Witten, hep-th/9612216.}
\lref\BST{E. Bergshoeff, E. Sezgin, P.K. Townsend,
``Supermembranes and eleven-dimensional supergravity,''
Phys. Lett. {\bf B189} (1987) 75;
``Properties of the eleven-dimensional super membrane theory,''
Annals Phys. {\bf 185} (1988) 330}
\lref\DHIS{M.J. Duff, P.S. Howe, T. Inami, K.S. Stelle,
``Superstrings in D = 10 from supermembranes in $D = 11$,''
Phys. Lett. {\bf 191B} (1987) 70.}
%
% forward equation references
%

\newsec{Introduction}

For many problems of D-brane physics,
it is useful to know the low energy action for
D-branes in curved space.  Given a solution of string theory, this
is defined by world-sheet computation along the lines of \refs{\dlp,\leigh}.
Such actions could also be used to define Matrix theory in curved space,
as discussed in \dstrings.

For small curvatures $\ap R << 1$,
a single D-brane in curved space is described by the
supersymmetrized Nambu-Born-Infeld Lagrangian.
At low energies, this reduces to two decoupled sectors,
$U(1)$ super Yang-Mills theory and a non-linear sigma model.

A crucial feature of the D-brane is enhanced gauge symmetry when
several D-branes coincide, and thus
an action describing more than one D-brane in curved space
must include a $U(N)$
super Yang-Mills theory as well as the non-linear sigma model,
but now the two sectors are coupled.

In \dcurve, a minimal set of axioms
was proposed, which should be satisfied both by weak coupling D-brane
actions in spaces of small curvature, and by Matrix theory actions.
We will review these below; they are rather uncontroversial
statements, the most novel being the
requirement that a string stretched between the
D-branes should have mass exactly proportional to the distance
between the branes.
It was then shown that a $d=4$, $\CN=1$ $U(N)$ Lagrangian with one chiral
superfield (parameterizing one curved complex dimension -- in other
words, describing a $3$-brane in six dimensions) was
determined (uniquely up to terms with more than two commutators)
by these axioms.  Dimensional reduction them produces actions
for all $p$-branes with $p<3$ in six dimensions.

Here we study the analogous problem of a $3$-brane in ten dimensions,
with three curved complex dimensions.
We will do a local analysis using normal coordinate expansions, and
develop the analogous ``D-normal coordinate expansion'' to sixth order.
We will show that the axioms cannot be satisfied unless the
target space is Ricci-flat, and give strong evidence that there exists
a solution when it is Ricci-flat.

Since the string theory definition works only in this case, this result
might seem natural and even predestined in the context of string theory.
However, since we did not use string theory to derive the action, but
rather a set of axioms which make perfect sense on a general curved
manifold, the result seems somewhat surprising.  We will discuss its
interpretation, as well as the possibility that this is a consistency
condition for Matrix theory compactifications, in the conclusions.

In section 2, we review the axioms. In section 3, we re-express these
axioms as conditions on the K\"ahler potential and the superpotential.
In section 4, we show that these conditions imply that the target
space must be Ricci flat, and that they
uniquely determine the superpotential.
We also find a compact expression for the conditions on the
K\"ahler potential. In section 5, by studying the normal coordinate
expansion to the sixth order, we demonstrate
that there exists a K\"ahler potential satisfying these conditions.
It is unique at the first non-trivial (fourth) order,
but at fifth order ambiguities begin to appear.
Section 6 is devoted to discussion.

\newsec{Axioms}

Given a $d$-dimensional K\"ahler manifold $\CM$ with K\"ahler potential $K_1$,
the problem will be to find
a $d=4$, $\CN=1$ $U(N)$ gauged non-linear sigma model satisfying
the axioms below.
The low energy action will be determined by a configuration space
$\CX_N$, a $dN^2$-dimensional K\"ahler manifold with potential $K_N$;
an action
of $U(N)$ by holomorphic isometries (which determines an action
of $GL(N)$); and a superpotential $W$.  The axioms are then

\item{1.}
The classical moduli space,
$$\{\CX_N|W'=0\}//U(N),$$
is the symmetric
product $\CM^N/S_N$.

\item{2.}
The generic unbroken gauge symmetry is $U(1)^N$, while if
two branes coincide the unbroken symmetry is $U(2)\times U(1)^{N-2}$,
and so on.

\item{3.}
Given two non-coincident branes at points $p_i\ne p_j$,
all states charged under $U(1)_i\times U(1)_j$ have mass
$m_{ij} = d(p_i,p_j)$, the distance along the shortest geodesic between
the two points.

\item{4.}
The action is a single trace (in terms of matrix coordinates),
\eqn\singletrace{
S = \Tr (\cdots) .
}

Some comments:

\item{(i)}
The axioms could have been stated in a coordinate-free
way.  Only axiom 4 used coordinates, and it could be replaced by
something like\nl
{\narrower
{{\narrower 4b.
The action has no explicit $N$ dependence, and its value (on field
configurations with no explicit $N$ dependence) is $O(N)$.}
}}\nl
We will give a coordinate-free statement of the problem elsewhere,
but the point is that we conjecture that the axioms imply the following
statement:\nl
Given local coordinates $z^i$ on $\CM$, there exist local matrix coordinates
$Z^i$ on $\CX_N$ (we define $\bar Z^i = (Z^i)^\dagger$),
such that the moduli space will be parameterized by
diagonal matrices whose eigenvalues are coordinates on the individual copies
of $\CM$, and such that the $U(N)$ action is
\eqn\unaction{
Z^i \rightarrow U^\dagger Z^i U .
}
Note that in choosing the coordinates $Z^i$, we must define the off-diagonal
matrix elements as coordinates on $\CX_N$, and \unaction\ is one constraint
on them.  It does not completely specify them, however: we can still make
holomorphic matrix changes of coordinate which are trivial on the moduli
space:
\eqn\holocoor{\eqalign{
Z^{i} & \longrightarrow
W^{i} = Z^{i}
+ a^{i}_{jk} [Z^{j},Z^{k}]
+ b^{i}_{jkl} [Z^{j},Z^{k}] Z^{l} + \cdots
\cr
\bar{Z}^{\bar{i}} &\longrightarrow
\bar{W}^{\bar{i}} = \bar{Z}^{\bar{i}}
+ \bar{a}^{\bar{i}}_{\bar{j}\bar{k}} [\bar{Z}^{\bar{k}}, \bar{Z}^{\bar{j}}]
+ \bar{b}^{\bar{i}}_{\bar{j}\bar{k}\bar{l}} \bar{Z}^{\bar{l}}
[\bar{Z}^{\bar{k}},\bar{Z}^{\bar{j}}]
+ \cdots
}}
where $ \bar{a}^{\bar{i}}_{\bar{j}\bar{k}} = (a^{i}_{jk} )^{*} $, etc. are
arbitrary constants.
\nl
Thus the problem will be to find all Lagrangians of the form
\eqn\matact{
\CL = \int d^4\theta \Tr K_N(Z,\bar Z) + \int d^2\theta \Tr \CW(Z) + \cc
}
satisfying the axioms, where we consider two Lagrangians equivalent
if they are related by a field redefinition \holocoor.

\item{(ii)}
The potential will be the sum of that coming from the D and F-terms,
and supersymmetric vacua will satisfy the conditions
\eqn\susyvac{\eqalign{
0 &= \sum_i [Z^i,{\p\ \Tr K_N\over \p Z^i}] \cr
0 &= {\p\ \Tr \CW_N\over \p Z_i}.
}}
We want the moduli space to be commuting matrices $[Z^i,Z^j]=0$, and
in more than one dimension the D-flatness condition alone does not
suffice to do this.
\nl
In flat space the model has the $\CN=4$, $d=4$ Lagrangian, which in this
notation has the superpotential $\Tr Z^1[Z^2,Z^3]$.  More generally, we
could take the form
\eqn\supform{
\CW(Z) = \sum_i \eps_{ijk} a^i(Z) [Z^j,Z^k]
}
which vanishes for any commuting matrices and thus has $\CW'=0$ on
this subspace.  For a generic function of this form, other solutions
would not be expected and thus we would have
the moduli space $\CM^N/S_N$.

\item{(iii)}
Clearly getting the right metric on moduli space will require us
to take for $K_N(Z,\bZ)$ some version of $K(z,\bz)$, with a definite
ordering prescription.  On the moduli space, the ordering
prescription will translate into a specific dependence of the action
for the off-diagonal elements on the point in moduli space.
\nl
The axioms stated can only determine the action and its second derivatives
on the moduli space, since they only refer to masses of stretched strings,
not interactions between stretched strings.
\nl
The second derivatives we will use are
\eqn\defhatg{
\hat g_{i\bj}(z_1,\bz_1,z_2,\bz_2) \equiv
 {\p^2 \over \p Z^i_{12} \p \bZ_{21}^\bj}
\Tr K_N(Z,\bZ)\bigg|_{Z\in \CM^N/S_N}
}
and
\eqn\defhatomega{
\hat \Omega_{ijk}(z_1,z_2) (z_1-z_2)^k \equiv
 {\p^2 \over \p Z^i_{12} \p Z_{21}^j}
\Tr \CW_N(Z)\bigg|_{Z\in \CM^N/S_N} .
}

\item{(iv)}
We could have considered a more general gauge kinetic term,
\eqn\gaugekin{
\Re \int d^2\theta\ \Tr f(Z)W^2 + f_{ij}(Z) [W,Z^i][W,Z^j].
}
Taking non-constant $f(Z)$ would lead to a space-dependent gauge coupling,
as would come from a non-constant dilaton background.
The mass conditions in this case involve the dilaton 
as well as the metric (e.g. see \sen),
and we will not consider this case here.
Non-constant $f_{ij}(Z)$, as we will see below, turns out to
be incompatible with the mass conditions.

\newsec{Mass conditions on K\"ahler potential and superpotential}

{}In this section, we summarize conditions on the K\"ahler
potential and superpotential imposed by the axioms. In particular,
we express the mass condition (the axiom 3) as conditions on
$\hat{g}(z_1,z_2)$ and $\hat{\Omega}_{ijk}(z_1,z_2)$ defined in
\defhatg\ and \defhatomega.

Our considerations will be in a neighborhood of a point $p$
in the target space,
and we will use coordinates $z$ in which this is at $z=0$.

\subsec{Normal coordinates and the squared distance}

On any K\"ahler manifold $\CM$,
one can always find a holomorphic
local coordinate system (normal coordinates) such that the metric has an
expansion around $z=0$ as
\eqn\gexpansion{\eqalign{
   g_{i\bj}(z,\bz) & = \delta_{i\bj} +
   \sum_{p=1}^{\infty} \sum_{q=1}^{\infty}
   g^{(p,q)}_{i \bj ; k_{1} \ldots k_{p} ; \bl_{1} \ldots \bl_{q}}
   z^{k_{1}} \cdots z^{k_{p}} \bz^{\bl_{1}} \cdots  \bz^{\bl_{q}}.
}}
By using K\"ahler potential transformations
$K_1\rightarrow K_1+\Re F(z)$, we can assume
the K\"ahler potential on $\CM$ has an expansion
\eqn\koneexpan{
   K_1(z,\bar z) = |z|^2 +
   \sum_{p=2}^{\infty} \sum_{q=2}^{\infty}
   K^{(p,q)}_{1\;\; ; k_{1} \ldots k_{p} ;\bl_{1} \ldots \bl_{q}}
    z^{k_{1}} \cdots z^{k_{p}} \bz^{\bl_{1}} \cdots  \bz^{\bl_{q}}.
}
More explicitly,
\eqn\konef{
K_1(z,\bz) = z^i \bz^i - {1\over 4}R_{i\bj k\bl} z^i\bz^\bj z^k\bz^\bl
+ \ldots .
}
where $R_{i\bj k\bl}$ is the Riemann curvature at $z=0$.
All symmetries of the Riemann tensor
$R_{i\bj k\bl}=R_{k\bj i\bl}$ and
$R_{i\bj k\bl}=R_{i\bl k\bj}$,
are manifest in this expression.

We also need an expression for the geodesic distance.
Let $d^2(x,z)$ be the squared geodesic distance from $x$ to $z$.
It obeys the differential equation
\eqn\deqn{
g^{i\bj}(x) {\p~ d^2(x,z)\over\p x^i}{\p~ d^2(x,z)\over\p \bx^\bj}
= d^2(x,z)
}
which follows from the Einstein relation 
$4g^{i\bar{j}} p_i p_{\bar{j}} = m^2$,
the momentum $p_i = m {\p S(x,y)/\p {x^i}}$
and the fact that the point particle action
is proportional to the geodesic distance, $S(x,y) = m d(x,y)$.

This combined with the initial condition
\eqn\sqrelone{
{\p^2 d^2(x,z)\over \p x^i \p \bx^\bj} \bigg|_{z=x} = g_{i\bj}(x),
}
determines the geodesic distance uniquely \hadamard.
For example, its expansion to $O(z^4)$ is
\eqn\dsqexp{\eqalign{
d^2(x,z) = |x-z|^2
&- {1\over 4} R_{i\bj k\bl} (x-z)^i (\bx-\bz)^\bj (x+z)^k (\bx+\bz)^\bl \cr
&- {1\over 12} R_{i\bj k\bl} (x-z)^i (\bx-\bz)^\bj (x-z)^k (\bx-\bz)^\bl
+ \ldots .
}}
In section 6, the expansion to the sixth order is given.

According to axiom 4, the K\"ahler
potential $K_N$ for the $N$ D-brane
is expressed as a trace of a power series in $Z$ and $\bZ$.
Again, by a holomorphic coordinate change $Z \rightarrow f(Z)$,
we can go to normal coordinates on the full configuration space
$\CX_N$, in which $K_N$ takes the form
\eqn\knexpan{
   \Tr K_N(Z,\bZ) = \Tr Z^i \bZ^i +
   \sum_{p,q=2}^{\infty}
   K_{N}^{(p,q)} \; \Tr
   Z \cdots \bZ \cdots Z \cdots \bZ \cdots.
}
Here $Z \cdots \bZ \cdots Z \cdots \bZ \cdots$ is some sequence
of $p$ $Z$'s and $q$ $\bZ$'s.
This eliminates the ambiguity expressed in equation \holocoor.

As discussed in \dcurve, reproducing the metric
for each of the $N$ branes requires
\eqn\modulimet{
    \Tr K_{N}(Z, \bZ)\bigg|_{Z\in \CM^N/S_N}=
    \sum_{a=1}^{N} K_{1}(z_{a},\bz_{a}),
}
where $z^i_1, \ldots, z^i_{N}$ are the eigenvalues of $Z^i$.
Thus $K_{N}$ must have the same expansion in powers of $Z$ and $\bZ$
as $K_{1}$, but the precise ordering remains to be
fixed.

A natural guess for the full K\"ahler form would be
\eqn\dkone{\eqalign{
   \Tr K_N(Z,\bZ) &= \Str K_1(Z,\bZ) \cr
   &= \Tr Z^i\bar Z^{\bar i}
   - {1\over 4} R_{i\bj k\bl}\; \STr Z^i \bZ^\bj Z^k \bZ^\bl
   + \cdots
}}
where $\Str$ is the symmetrized trace:
$\Str A_1 \cdots A_{n}={1\over n!}\sum_{\sigma \in S_{N}}
\Tr A_{\sigma(1)} \cdots A_{\sigma(n)}$.
However, there could be additional terms
which vanish on the moduli space, and we will need to constrain them
by using the mass condition (axiom 3).
As it will turn out, \dkone\ must be corrected (at sixth order!).

\subsec{The mass condition for gauge bosons}

The simplest thing to check is that the gauge boson masses
are correctly reproduced, because these do not depend on the superpotential.
This requires
\eqn\Higgsmass{
   d^2(z_1,z_2) = {1\over 2} 
   {d^2\over d t d\bar t} \Tr K_{N} (Z + t[X,Z], \bZ +\bar t[X,\bZ])
   \bigg|_{Z\in \CM^N/S_N, \;  t=0}
}
where $X=E_{12}+E_{21}$ is a broken generator.
More explicitly,
\eqn\masscontwo{
\hat g_{i\bj}(z_1,\bz_1,z_2,\bz_2) (z_1-z_2)^i (\bz_1-\bz_2)^\bj
 = d^2(z_1,z_2) .
}

\subsec{Non-constant gauge kinetic term}

We now consider the case of non-constant $f_{ij}(z)$ in \gaugekin.
This would modify the mass condition to
\eqn\masscontwogen{
\hat g_{i\bj}(z_1,\bz_1,z_2,\bz_2) (z_1-z_2)^i (\bz_1-\bz_2)^\bj
 = d^2(z_1,z_2) \Re f(z_1,z_2) 
}
where $f(z_1,z_2)$ is holomorphic.

Suppose $f(z_1,z_2) \sim 1 + z^k$ is non-constant;
then the left hand side will include a term $O(z^{k+1} \bz) + \cc$.
{}From \knexpan, we see that in normal coordinates
$\hat g_{i\bj} \sim \delta_{i\bj} + O(z \bz)$, leading to the
right hand side $|z|^2 + O(z^2\bz^2)$ which cannot include such a term.

Thus the mass condition requires trivial gauge kinetic term,
and we restrict attention to this case.

\subsec{The mass condition for scalars}

The masses of the chiral superfields are determined by the kinetic
term $\hat g_{i\bj}$ and the second
derivatives of the potential, and thus their mass condition is
\eqn\threeD{
    \delta_{i} \bar{\delta}_{{\bj}} V
    =
    \delta_{i} \bar{\delta}_{{\bj}} V_{D} +
    \delta_{i} \bar{\delta}_{{\bj}} V_{F}
    \equiv d^2 \hat{g}_{i\bj},
}
where $\delta_{i} = {\p / \p Z^{i}_{12}}$ and 
$\bar{\delta}_{\bj} = {\p / \p \bZ^{\bj}_{21}}$.

The D-term contribution is
\eqn\dtermpot{
V_D = -\half\sum_{i,\bar j} \Tr [Z^i,{\p K\over \p Z^i}]
        [\bZ^\bj,{\p K\over \p \bZ^\bj}].
}
and its second derivative on the moduli space was given in \dcurve,
while the $F$-term contribution is
\eqn\ftermpot{
  V_{F} =
  \left(
  {\del^2 K \over \del Z_{ab}^{i} \del \bZ_{ba}^{{\bj}}}
  \right)^{-1}
  {\del W \over \del Z_{ab}^{i}}
  {\del \bar{W} \over \del \bZ_{ba}^{{\bj}}}.
}
with second derivative on the moduli space
\eqn\deldelVC{\eqalign{
  {\del^2 V_{F} \over \del Z_{12}^{i} \del \bZ_{21}^{{\bj}} }  & =
  \left(
  {\del^2 K \over \del Z_{12}^{k} \del \bZ_{21}^{{\bm}}}
  \right)^{-1}
  {\del^2 W \over \del Z_{12}^{i} \del Z_{21}^{k}}
  {\del^2 \bar{W} \over \del \bZ_{21}^{{\bj}} \del Z_{12}^{{\bm}}} \cr
&
  = \hat{g}^{k{\bm}}
  \hat\Omega_{ikl}
  (z_{1}^{l} - z_{2}^{l})
  \hat\Omega_{{\bj}\bm\bn}
  (\bz_{1}^{\bn} - \bz_{2}^{\bn})
}}
More explicitly, with the gauge boson mass \masscontwo,
the mass condition \threeD\ reads
\eqn\masscond{\eqalign{
    &
    \hat{g}_{i\bn}
    (\bz^{\bn}_{1} - \bz^{\bn}_{2})
    \hat{g}_{m\bj}
    (z^{m}_{1} - z^{m}_{2})
    +
    \hat{g}^{k \bl}
    \hat\Omega_{ikm}
    (z_{1}^{m} - z_{2}^{m})
    \hat\Omega_{\bj\bl\bn}
    (\bz_{1}^{\bn} - \bz_{2}^{\bn})
    \cr
    & \qquad
    \equiv
    \hat{g}_{i\bj} \hat{g}_{m\bn}
    (z_{1}^{m} - z_{2}^{m})
    (\bz_{1}^{\bn} - \bz_{2}^{\bn}).
}}

\newsec{More on the mass conditions}

In this section, we study the mass conditions \masscontwo\ and
\masscond\ found in the previous section. We show that these
require that the target space to be Ricci-flat and also determine
the superpotential completely.

\subsec{Ricci flatness}

In this subsection, we will show that 
the mass condition \masscond\ implies $R_{i\bj}=0$.

Note that the off-diagonal metric $\hat g$ and diagonal metric
$g$ are related as
\eqn\gandghat{
  \lim_{z_{1}\rightarrow z_{2}}
  \hat{g}_{i\bj}(z_{1},\bz_{1},z_{2},\bz_{2})
  =g_{i\bj}(z_{2},\bz_{2}),
}
which follows from \sqrelone\  and \masscontwo.
For later convenience, set
$$
   \Omega_{ijk}(z_{2})= \lim_{z_{1} \rightarrow z_{2}}
   {\p \over \p z_{1}^{k}}
   \left\{
   \hat\Omega_{ijm}(z_{1},z_{2}) (z_{1}-z_{2})^{m}
   \right\}.
$$
Taking
$\lim_{z_{1} \rightarrow z_{2}} {\p^2\over \p z^{m}_{1}\p
\bz^{\bn}_{2}}$
of \masscond, we get
\eqn\masscondlimit{
    g_{i\bn} g_{m\bj}
    +
    g^{k \bl}
    \Omega_{ikm}
    \Omega_{\bj\bl\bn}
    =
    g_{i\bj} g_{m\bn}.
}

As was explained in the introduction, we assume
a superpotential of the form
\eqn\WNgeneral{
   \CW_{N}(Z)= c \, \Tr Z^{1} [Z^2, Z^3]
    + \cdots
}
where $\cdots$ stands for higher order terms with at least one
commutator. Then, $\Omega_{ijk}(z) $ can be expanded as
\eqn\omegaexpansion{\eqalign{
\Omega_{ijk}(z)  & = c \, \epsilon_{ijk} +
  \sum_{p \geq 1}
  \Omega^{(p)}_{ijk;m_{1} \ldots m_{p}} z^{m_{1}} \cdots z^{m_{p}}
}}
Let us plug the expansions \gexpansion\ and \omegaexpansion\
into \masscondlimit.
On the right hand side, there is no purely holomorphic
terms in the expansion.
On the left hand side, however,
such terms would appear with coefficients
$ \delta^{k \bl} \Omega^{(p)}_{ikm;m_1\cdots m_p} \epsilon_{\bj\bl\bn}.$
This implies
$\Omega^{(p)}_{ikm;m_1\cdots m_p} = 0 $ for all $p\geq 1$, i.e.
$\Omega_{ijk}(z)$ actually has no $z$ dependence in our coordinate system;
\eqn\OmegaEps{
\Omega_{ijk}(z) =  c \, \epsilon_{ijk}.
}
Then, \masscondlimit\  reduces to
$$
  g_{i\bn} g_{m\bj} +
  |c|^2 g^{k \bl} \epsilon_{ikm} \epsilon_{\bj\bl\bn}
  = g_{i\bj} g_{m\bn},
$$
which is equivalent to $\det g_{m\bn}(z,\bz) = |c|^2$.
The normalization \gexpansion\ fixes $|c|=1$. Thus the final result is
\eqn\detgone{
  \det g_{m\bn}(z,\bz) \equiv 1.
}
Ricci flatness of $\CM$ is a corollary of this:
\eqn\RFasCor{
   R_{i\bj} \equiv {\p^2 \over \p z^{i} \p \bz^{\bj}} \log \det
   g_{m\bn}(z,\bz)
   = 0.
}

\subsec{Superpotential $\CW_{N}(Z)$}

In this subsection we will show that the mass condition
\masscond\ determines the superpotential $\Tr \CW_{N}(Z)$ to be
\eqn\superpotW{
  \Tr \CW_{N}(Z) = \Tr Z^1 [Z^2, Z^3],
}
to all order in diagonal coordinates but up to the
second order in commutators.

The basic strategy is quite similar to that used in the previous subsection.
Here, instead of taking the limit $z_1 \rightarrow z_2$, we will expand
various quantities as a power series in $z_{1}-z_{2}$ and
$\bz_{1}-\bz_{2}$:
\eqn\Aexpansion{\eqalign{
    A(z_{1},\bz_{1},z_{2},\bz_{2}) =
    \sum_{p,q=0}^{\infty} &
    A^{(p,q)}_{i_1 \cdots i_p;\bj_1 \cdots \bj_q}
    \left(
    {\textstyle {z_1+z_2 \over 2}, {\bz_1+\bz_2 \over 2}}
    \right) \cr
    & \qquad \times
    (z_1-z_2)^{i_1} \cdots
    (z_1-z_2)^{i_p}
    (\bz_1-z_2)^{\bj_1} \cdots
    (\bz_1-z_2)^{\bj_q}.
}}

The second derivative of the superpotential is expanded as
\eqn\omegahat{\eqalign{
   \hat{\Omega}_{ijk}(z_{1}-z_{2})^k
& =
   {\p^2 \over \p Z_{12}^{i}  \p Z_{21}^{j}}
   \Tr \CW_{N}(Z)
     \bigg|_{Z\in \CM^N/S_N} \cr
& =
   \sum_{p=1}^{\infty}
     \hat{\Omega}^{(p)}_{ij;k_1\ldots k_{p}}
     \left(\textstyle{z_1+z_2 \over 2} \right)
     (z_1-z_2)^{k_1} \cdots (z_1-z_2)^{k_p}
}}
{}From the superpotential \WNgeneral, we have
$$
  \hat{\Omega}^{(1)}_{ij;k}\left(\textstyle{z_1+z_2 \over 2} \right)
   = c \epsilon_{ijk}
  + O\left(\textstyle {z_1 + z_2 \over 2}\right).
$$
We saw in the previous subsection that $|c|=1$ and by choice of
coordinates we can take $c=1$. Then, \OmegaEps\ implies
$$
  \Omega_{ijk}(z_2)=
  \lim_{z_1 \rightarrow z_2}
   {\p \over \p z_{1}^{k}}
   \left\{
   \hat\Omega_{ijm}(z_{1},z_{2}) (z_{1}-z_{2})^{m}
   \right\}
  = \Omega^{(1)}_{ij;k}(z_2)
  = \epsilon_{ijk}.
$$
$\delta_{i} \delta_{j} \Tr \CW_{N}$ now reads
\eqn\omegahattwo{\eqalign{
   \hat{\Omega}_{ijk}(z_{1}-z_{2})^k
   = & \epsilon_{ijk}
     + \sum_{p=2}^{\infty}
     \hat{\Omega}^{(p)}_{ij;k_1\ldots k_{p}}
     \left(\textstyle{z_1+z_2 \over 2} \right)
     (z_1-z_2)^{k_1} \cdots (z_1-z_2)^{k_p}.
}}
On the other hand, the D-K\"ahler potential \knexpan\
gives the following expansion of $\hat{g}$
\eqn\ghatexpansion{\eqalign{
   \hat{g}_{i\bj}(z_{1},\bz_{1},z_{2},\bz_{2}) = \delta_{i\bj} +
   & \sum_{p=1}^{\infty} \sum_{q=1}^{\infty}
   \hat{g}^{(p,q)}_{i \bj ; k_{1} \ldots k_{p} ; \bl_{1} \ldots \bl_{q}}
     \left(\textstyle{z_1+z_2 \over 2},{\bz_1+\bz_2 \over 2} \right) \cr
   & \quad \times
     (z_1-z_2)^{k_1} \cdots (z_1-z_2)^{k_p}
     (\bz_1-z_2)^{\bl_1} \cdots (\bz_1-z_2)^{\bl_q}.
}}

Let us consider a similar expansion of the mass degeneracy condition
\masscond. On the right hand side, there is no terms with bi-degree
$(p=1,q\geq 2)$. On the left hand side, however,
such terms would appear with coefficients
$ \delta^{k \bl} \epsilon_{\bj\bl\bn} \Omega^{(p)}_{ik;m_1\cdots m_p}
$.
Thus, $\hat{\Omega}^{(p)}$ must vanish for all $p\geq 2$.
This implies
\eqn\Omegaz{
   \hat{\Omega}_{ijk}(z_{1},z_{2}) \, (z_{1}-z_{2})^k
   = \epsilon_{ijk} (z_{1}-z_{2})^k,
}
to all orders in $z_1$ and $z_2$, and the stated result.

\subsec{Condition on $\det\hat{g}$}

Given the $F$ term contribution \Omegaz,
we can reduce the mass condition \masscond\  to a much simpler form:
\eqn\newmasscond{\eqalign{
&
\hat{g}_{i\bn}\hat{g}_{m\bj}
  (z^{m}_{1} - z^{m}_{2})
  (\bz^{\bn}_{1} - \bz^{\bn}_{2})
+
\hat{g}^{k \bl}
  \epsilon_{ikm}
  \epsilon_{\bj\bl\bn}
  (z_{1}^{m} - z_{2}^{m})
  (\bz_{1}^{\bn} - \bz_{2}^{\bn})
\cr
& \qquad
  \equiv
\hat{g}_{i\bj} \hat{g}_{m\bn}
  (z_{1}^{m} - z_{2}^{m})
  (\bz_{1}^{\bn} - \bz_{2}^{\bn}).
}}
which can be rewritten as
\eqn\newnewmasscond{
(1 - \det \hat{g}) \times
\hat{g}^{k \bl}
  \epsilon_{ikm}
  \epsilon_{\bj\bl\bn}
  (z_{1}^{m} - z_{2}^{m})
  (\bz_{1}^{\bn} - \bz_{2}^{\bn}) = 0.
}
This holds if and only if
\eqn\detghatone{
   \det \hat{g}(z_1,\bz_1,z_2,\bz_2) \equiv 1.
}
The condition \detghatone\ bears a striking similarity
to the previous result
$\det g(z,\bz) \equiv 1.$
This is consistent since $g$ and $\hat{g}$ are
related via \gandghat.

\newsec{Explicit form of D-K\"ahler potential}

We have seen that the axioms of D-geometry and especially
the mass condition \masscond\ puts stringent constraints
on our problem. In normal coordinates,
the superpotential $\CW_{N}(Z)$ is uniquely
fixed as \superpotW\ and the nontrivial information of D-geometry
is encoded in the K\"ahler potential $K_{N}(Z,\bZ)$.

Let us summarize the properties $K_{N}(Z,\bZ)$ should have:

\item{1.}
It must reproduce the Ricci flat metric for each of the $N$ branes.
\eqn\modulimet{
\Tr K_{N}(Z, \bZ)\bigg|_{Z\in \CM^N/S_N}=
\sum_{a=1}^{N} K_{1}(z_{a},\bz_{a}),
}
where $z^i_1, \ldots, z^i_{N}$ are the eigenvalues of $Z^i$.

\item{2.}
The gauge boson mass condition: Its second derivative
must reproduce the geometric distance.
\eqn\geoddist{
   \hat g_{i\bj}(z_1,\bz_1,z_2,\bz_2) (z_1-z_2)^i (\bz_1-\bz_2)^\bj
   = d^2(z_1,z_2) .
}

\item{3.} In normal coordinates, the general mass condition becomes
\eqn\detcond{
   \det \hat{g}_{i\bj}(z_1,\bz_1,z_2,\bz_2) \equiv 1.
}

\subsec{$K_N$ to the fourth order}

We now show that, at the first non-trivial
order (fourth order) in the normal coordinate expansion,
there exists a unique $K_N$ satisfying these conditions:
\eqn\kfinal{
   K_N(Z,\bZ) = \Tr Z^i \bZ^{\bar{i}} -
   {1\over 4} R_{i\bj k\bl} \STr Z^i Z^k \bZ^\bj \bZ^\bl .
}

The most general form of $K_N$ to fourth order reads
\eqn\ktwoD{\eqalign{
    \Tr K_N(Z,\bZ) = \Tr Z\bZ
    &- {1\over 4} R_{i\bj k\bl} \STr Z^i Z^k \bZ^\bj \bZ^\bl \cr
    &+ {1\over 4}A_{(ij)[\bk\bl]} \Tr \{Z^i,Z^j\}[\bZ^\bk,\bZ^\bl] \cr
    &+ {1\over 4}A^*_{[ij](\bk\bl)} \Tr [Z^i,Z^j]\{\bZ^\bk,\bZ^\bl\} \cr
    &+ {1\over 4}B_{[ij][\bk\bl]} \Tr [Z^i,Z^j][\bZ^\bk,\bZ^\bl] \cr
    &+ {1\over 2}C_{(ik)(\bj\bl)} \Tr [Z^i,\bZ^\bj][Z^k,\bZ^\bl] \cr
    &+ D_{[ik][\bj\bl]} \Tr Z^i\bZ^\bj Z^k\bZ^\bl \cr
}}

In general, the terms $B$ and $C$ together are redundant,
by using the Jacobi identity.
For example, $\tr [x,\bx][y,\by] - \tr [x,\by][y,\bx] = \tr [x,y][\bx,\by]$.
So, we require that $C$ be symmetric under such interchanges.
Similarly, the symmetry condition on $D$ makes it
independent of $C$ and $R$.

To check that this includes all fourth order terms, we count
$d^4$ terms $\Tr Z^i Z^j \bZ^k \bZ^l$ and $d^2(d^2+1)/2$ of the form
$\Tr Z^i \bZ^\bj Z^k \bZ^\bl$, while the expansion above has
$(d(d+1)/2)^2 + 2(d(d-1)/2)^2 +2(d(d-1)/2)(d(d+1)/2) + (d(d+1)/2)^2$ terms,
which agrees.

To check the conditions at fourth order we use \dsqexp\ for $d^2$, and compute
\defhatg\ using \ktwoD\ to find
\eqn\ghatfour{\eqalign{
    \hat g_{i\bj}(z_1,\bz_1,z_2,\bz_2) =
    \delta_{i\bj}
    &- {1\over 4} R_{i\bj k\bl} (z_1+z_2)^k (\bz_1+\bz_2)^\bl \cr
    &- {1\over 12} R_{i\bj k\bl} (z_1-z_2)^k (\bz_1-\bz_2)^\bl \cr
    &+ (A_{ik \bj\bl}+D_{ik \bj\bl}) (z_1+z_2)^k (\bz_1-\bz_2)^\bl \cr
    &+ (A^*_{ik \bj\bl}+D_{ik \bj\bl}) (z_1-z_2)^k (\bz_1+\bz_2)^\bl \cr
    &+ B_{ik \bj\bl} (z_1-z_2)^k (\bz_1-\bz_2)^\bl \cr
    &- C_{ik \bl\bj} (z_1-z_2)^k (\bz_1-\bz_2)^\bl .
}}
We find that, taking into account the symmetries of the tensors,
we need $A=C=D=0$ but $B$ is undetermined.

Now let us use condition 3.
Using our computation \ghatfour\ of $\hat g$,
we have (to second order)
\eqn\ghatdet{
\det \hat g_{i\bj} = 1
- {1\over 4} R_{k\bl} (z_1+z_2)^k (\bz_1+\bz_2)^\bl
- ({1\over 12} R_{k\bl}-B_{k\bl}) (z_1-z_2)^k (\bz_1-\bz_2)^\bl,
}
where $R_{k\bl} = \delta^{i\bj} R_{i\bj k\bl}$ is the Ricci tensor
at $z=0$ and $B_{k\bl} = \delta^{i\bj} B_{ik\bj\bl}$.
We see that in this
coordinate system the only solutions have
$R_{k\bl}=B_{k\bl}=0$. The latter implies $B_{[ik][\bj\bl]}=0$
in $d\leq 3$. 

\subsec{Higher orders}

The appearance of Ricci flatness in the computation was somewhat unexpected,
and to get more insight (and be sure that there are no further consistency
conditions) we push the computation to the next non-trivial case, which
is sixth order.
In addition, we would like to know whether the uniqueness we found
at fourth order persists at higher orders.

\subsec{Normal coordinate expansions to sixth order}

To this order, the expansion of the K\"ahler potential in normal
coordinates is
\eqn\ktosixth{\eqalign{
    K_0(z,\bz) = &
    z^i \bz^{\bi} + \sum_{p,q=2}^{\infty}
    %\sum_{i_{1}, \ldots, i_{p}=1}^d
    %\sum_{\bj_{1}, \ldots, \bj_{q}=1}^d
    K^{(p,q)}_{i_{1}, \ldots, i_{p};\bj_{1},\ldots,\bj_{q}}
    z^{i_{1}} \cdots z^{i_{p}}
    \bz^{\bj_{1}} \cdots \bz^{\bj_{q}}
    \cr
    = & z^i \bz^i - {1\over 4}R_{i\bj k\bl} z^i\bz^\bj z^k\bz^\bl \cr
    & + S_{i\bj k\bl m} z^i\bz^\bj z^k\bz^\bl z^m  + \cc \cr
    & + T_{i\bj k\bl m p } z^i\bz^\bj z^k\bz^\bl z^m z^p + \cc \cr
    & + U_{i\bj k\bl m \bn} z^i\bz^\bj z^k\bz^\bl z^m \bz^\bn \cr
    & + \cdots
}}
where the coefficients $S$, $T$ and $U$ are given by
\eqn\STU{\eqalign{
    %R_{i\bj k \bl} & = R_{i\bj k \bl}(0) \cr
    S_{i\bj k\bl m} & = -{1\over 12} {\p \over \p z^{m}} R_{i\bj k \bl}(z)
    \biggr|_{z=0}
    \cr
    T_{i\bj k\bl m p} & = -{1\over 48} {\p^2 \over \p z^{m} \p z^{p}}
    R_{i\bj k \bl}(z)\biggr|_{z=0}
    \cr
    U_{i\bj k\bl m \bn} & = - {1\over 36 } \left(
    {\p^2 \over \p z^{m} \p \bz^{\bn}}
    R_{i\bj k \bl}(z)\biggr|_{z=0}
    - \delta^{p\bq} R_{i \bar{q} k \bn}(0) R_{p\bj m \bl}(0)
    \right)
}}
This can be checked by using the expression
\eqn\curvature{
    R_{i\bj k \bl}(z)  =
    -
    {\p^4 K\over \p z^i \p \bz^\bj \p z^k \p \bz^\bl }
    +
    g^{m \bn }
    {\p^3 K\over \p z^i \p z^k \p \bz^\bn }
    {\p^3 K\over \p z^m \p \bz^\bj \p \bz^\bl } .
}

The terms $S$ and $T$ correspond to purely holomorphic derivatives
of the curvature and as such do not lead to essentially new features.
However, the mixed derivatives $U z^3 \bz^3$ might.

Again by using \sqrelone\ and \deqn,
we find the expansion of the
squared geodesic distance $d^2(x,z)$
to $O(z,\bz)^6$:
\eqn\dsqexp{\eqalign{
    d^2(x,z) = & |x-z|^2 \cr
    &- {1\over 4} R_{i\bj k\bl} (x-z)^i (\bx-\bz)^\bj (x+z)^k (\bx+\bz)^\bl \cr
    &- {1\over 12} R_{i\bj k\bl} (x-z)^i (\bx-\bz)^\bj (x-z)^k
    (\bx-\bz)^\bl \cr
    &
    + {3\over 4} S_{i\bj k\bl m}
    (x-z)^i (\bx-\bz)^\bj (x+z)^k (\bx+\bz)^\bl (x+z)^m
    \cr
    &
    + {1\over 4} S_{i\bj k\bl m}
    (x-z)^i (\bx-\bz)^\bj (x-z)^k (\bx+\bz)^\bl (x-z)^m
    \cr
    &
    + {1\over 2} S_{i\bj k\bl m}
    (x-z)^i (\bx-\bz)^\bj (x-z)^k (\bx-\bz)^\bl (x+z)^m
    \cr
    &
    + \cc \cr
    &
    + {1\over 10} T_{i\bj k\bl m p}
    (x-z)^i (\bx-\bz)^\bj (x-z)^k (\bx-\bz)^\bl (x-z)^m (x-z)^p
    \cr
    &
    + {1\over 2} T_{i\bj k\bl m p}
    (x-z)^i (\bx-\bz)^\bj (x+z)^k (\bx-\bz)^\bl (x+z)^m (x-z)^p
    \cr
    &
    + {1\over 2} T_{i\bj k\bl m p}
    (x-z)^i (\bx-\bz)^\bj (x+z)^k (\bx+\bz)^\bl (x-z)^m (x-z)^p
    \cr
    &
    + {1\over 2} T_{i\bj k\bl m p}
    (x-z)^i (\bx-\bz)^\bj (x+z)^k (\bx+\bz)^\bl (x+z)^m (x+z)^p
    \cr
    &
    + \cc \cr
    &
    + {9\over 80} U_{i\bj k\bl m \bn}
    (x-z)^i (\bx-\bz)^\bj (x-z)^k (\bx-\bz)^\bl (x-z)^m (\bx-\bz)^\bn
    \cr
    &
    + {3\over 16} U_{i\bj k\bl m \bn}
    (x-z)^i (\bx-\bz)^\bj (x+z)^k (\bx-\bz)^\bl (x+z)^m (\bx-\bz)^\bn
    \cr
    &
    + {3\over 4} U_{i\bj k\bl m \bn}
    (x-z)^i (\bx-\bz)^\bj (x+z)^k (\bx+\bz)^\bl (x-z)^m (\bx-\bz)^\bn
    \cr
    &
    + {3\over 16} U_{i\bj k\bl m \bn}
    (x-z)^i (\bx-\bz)^\bj (x-z)^k (\bx+\bz)^\bl (x-z)^m (\bx+\bz)^\bn
    \cr
    &
    + {9\over 16} U_{i\bj k\bl m \bn}
    (x-z)^i (\bx-\bz)^\bj (x+z)^k (\bx+\bz)^\bl (x+z)^m (\bx+\bz)^\bn
    \cr
    &
    - {1 \over 720}
    R_{i\bq k \bn} \delta^{p\bq} R_{p\bj m \bl}
    (x-z)^i (\bx-\bz)^\bj (x-z)^k (\bx-\bz)^\bl (x-z)^m (\bx-\bz)^\bn
    \cr
    &
    - {1 \over 48}
    R_{i\bq k \bn} \delta^{p\bq} R_{p\bj m \bl}
    (x-z)^i (\bx-\bz)^\bj (x-z)^k (\bx-\bz)^\bl (x+z)^m (\bx+\bz)^\bn
    \cr
    & + \cdots
}}

\vfill\eject

\subsec{D-K\"ahler potential to fifth order}

We now consider the possible fifth order
terms in the D-K\"ahler potential,
and their contributions to $\hat g_{i\bj}$.

The natural guess for the fifth order term is
\eqn\Kfifthord{
K^{(5)}(Z,\bZ) = \ldots +
 S_{i\bj k\bl m} \Str Z^i\bZ^\bj Z^k\bZ^\bl Z^m  + {\rm c.c.} + \ldots .
}

This leads to the variation
\eqn\ghatfive{\eqalign{
\hat{g}^{(5)}_{i\bar{j}}(x,z)   =
    &{3\over 4} S_{i\bj k\bl m} (x+z)^k (\bx+\bz)^\bl (x+z)^m
    \cr
    &+ {1\over 4} S_{i\bj k\bl m} (x-z)^k (\bx+\bz)^\bl (x-z)^m
    \cr
    &
    + {1\over 2} S_{i\bj k\bl m} (x-z)^k (\bx-\bz)^\bl (x+z)^m + {\rm c.c.}
}}
which indeed reproduces the corresponding terms of \dsqexp\ in
\geoddist.

However, it is not the unique K\"ahler potential which does so.
In fact, there are  two commutator terms one can add which 
modify $\hat g_{i\bj}$, but which preserve both \geoddist\ and \detcond\
at this order.
An explicit example is
\eqn\fifthamb{
K^{(5)}(Z,\bZ) = 
E_{[ij]k\bl\bm} \Tr [Z^i,Z^j] \{ [Z^k,\bZ^\bl],\bZ^\bm\} + 
E^*_{[\bi\bj]\bk lm} \Tr [\bZ^\bi,\bZ^\bj] \{ [\bZ^\bk,Z^l],Z^m\}.
}
This leads to the variation
\eqn\ghatfiveamb{\eqalign{
\hat{g}^{(5)}_{i\bar{j}}(x,z)   =
2 E_{ilk\bj\bm} (z-x)^l (x-z)^k (\bx+\bz)^\bm + 
2 E^*_{\bj\bl\bk im} (\bz-\bx)^\bl (\bx-\bz)^\bk (x+z)^m .
}}
The symmetry of $E$ guarantees that this will reproduce \geoddist,
while at this order, the condition $\det \hat g =1 $ reduces to
$\tr \hat g^{(5)} = 0$, which also has solutions, e.g.
$E_{ijk\bl\bm} = \eps_{ij} \delta_{k\bl} c_\bm$.

This means that the axioms as stated do not have a unique solution.
To clarify the degree of non-uniqueness,
let us check the Riemann curvature on the moduli space. One can show
that the mixed components are given by
\eqn\mixedR{
 R_{Z^i_{11}+Z^i_{22}\; \bZ^{\bar{j}}_{11}+\bZ^{\bar{j}}_{22}\;
 Z^{k}_{12} \; \bZ^{\bar l}_{21}} (z_{1},z_{2})
  = R_{i \bj k \bl} \, (\textstyle {z_1+z_2\over 2}) + 
    O(z_1 - z_2).
}
The ambiguities discussed above manifest themselves at 
$O(z_1 -z_2)$. 
In terms of the geometry of the configuration space $\CX_N$, 
the curvature on the bundle of off-diagonal modes is fixed
on the diagonal $z_i=z_j$, but is not completely fixed off the diagonal.

\subsec{D-K\"ahler potential to sixth order}

We give an explicit solution of the conditions to this order,
without studying the question of uniqueness.

The gauge boson mass condition
\masscontwo\ can be reproduced by the K\"ahler potential
\eqn\Ksixord{\eqalign{
    K(Z,\bZ) = &
    \tr Z\bZ \cr
    & - {1\over 4} R_{i\bj k\bl}~ \Str Z^i \bZ^\bj Z^k \bZ^\bl \cr
    & + S_{i\bj k\bl m} \Str Z^i\bZ^\bj Z^k\bZ^\bl Z^m  + \cc \cr
    & + T_{i\bj k\bl m p } \Str Z^i\bZ^\bj Z^k\bZ^\bl Z^m Z^p + \cc \cr
    & + U_{i\bj k\bl m \bn} \Str Z^i\bZ^\bj Z^k\bZ^\bl Z^m \bZ^\bn \cr
    & + R_{i\bq m \bn} \delta^{\bq p} R_{p\bl k\bj} \cr
    & \quad\times \biggl( 
        - {1\over 96} \, \tr 
        \{ Z^{k}    , [Z^{i},\bZ^{\bl}] \}
        \{ \bZ^{\bn}, [Z^{m},\bZ^{\bj}] \} \cr
    & \qquad\quad
        + {1\over 1440} \, \tr 
        [ Z^{k} , [Z^{i},\bZ^{\bl}] \, ] \,
        [ \bZ^{\bn}, [Z^{m},\bZ^{\bj}] \, ] \biggr) \cr
    & + F_{k i \bl m[\bn\bq]}
  \; \tr \{Z^k, [Z^i,\bZ^\bl]\}
       \{Z^m, [\bZ^\bn,\bZ^\bq]\} + \cc \cr
    & + G_{k i \bl m [\bn\bq]}
  \; \tr [Z^k, [Z^i,\bZ^\bl]\,]\,
       [Z^m, [\bZ^\bn,\bZ^\bq]\,] + \cc \cr
    & \cr
    = &
    \sum_{p,q \atop p+q\leq 6}
    K^{(p,q)}_{i_{1}, \ldots, i_{p};\bj_{1},\ldots,\bj_{q}}
    \Str
    Z^{i_{1}} \cdots Z^{i_{p}}
    \bZ^{\bj_{1}} \cdots \bZ^{\bj_{q}} \cr
    & + R_{i\bq m \bn} \delta^{\bq p} R_{p\bl k\bj} \cr
    & \quad\times \biggl( 
        - {1\over 96} \, \tr 
        \{ Z^{k}    , [Z^{i},\bZ^{\bl}] \}
        \{ \bZ^{\bn}, [Z^{m},\bZ^{\bj}] \} \cr
    & \qquad\quad
        + {1\over 1440} \, \tr 
        [ Z^{k} , [Z^{i},\bZ^{\bl}] \, ] \,
        [ \bZ^{\bn}, [Z^{m},\bZ^{\bj}] \, ] \biggr) \cr
    & + F_{k i \bl m[\bj\bn]}
  \; \tr \{Z^k, [Z^i,\bZ^\bl]\}
       \{Z^m, [\bZ^\bj,\bZ^\bn]\} + \cc \cr
    & + G_{k i \bl m [\bj\bn]}
  \; \tr [Z^k, [Z^i,\bZ^\bl]\,]\,
       [Z^m, [\bZ^\bj,\bZ^\bn]\,] + \cc \cr
}}

Note that it is not a symmetric trace.  The extra $O(R^2)$ terms
are required to satisfy \geoddist, while the terms $F$ and $G$ will
be required to satisfy \detcond.

This $K(Z,\bZ)$ gives $\hat{g}$ as
\eqn\ghatsix{\eqalign{
    \hat{g}_{i\bar{j}}(x,z)
    := &  {\p^2 \over \p Z_{12}^{i} \p \bZ_{21}^\bj}
    K \bigg|_{Z\in \CM^N/S_N} \cr
    = & \,
    \delta_{i\bar{j}} \cr
    &
    - {1\over 4} R_{i\bar{j}k\bar{l}}
    (x + z)^k (\bx + \bz)^\bl \cr
    &
    - {1\over 12} R_{i\bar{j}k\bar{l}}
    (x - z)^k (\bx - \bz)^\bl \cr
    &
    + {1\over 4} S_{i\bj k\bl m} (x-z)^k (\bx+\bz)^\bl (x-z)^m
    \cr
    &
    + {1\over 2} S_{i\bj k\bl m} (x-z)^k (\bx-\bz)^\bl (x+z)^m
    \cr
    &
    + {3\over 4} S_{i\bj k\bl m} (x+z)^k (\bx+\bz)^\bl (x+z)^m
    \cr
    &
    + \cc
    \cr
    &
    + {1\over 10} T_{i\bj k\bl m p }  (x-z)^k (\bx-\bz)^\bl (x-z)^m (x-z)^p
    \cr
    &
    + {1\over 2} T_{i\bj k\bl m p }  (x+z)^k (\bx+\bz)^\bl (x-z)^m (x-z)^p
    \cr
    &
    + {1\over 2} T_{i\bj k\bl m p }  (x-z)^k (\bx-\bz)^\bl (x+z)^m (x+z)^p
    \cr
    &
    + {1\over 2} T_{i\bj k\bl m p }  (x+z)^k (\bx+\bz)^\bl (x+z)^m (x+z)^p
    \cr
    &
    + \cc
    \cr
    &
    + \left( {9\over 80} U_{i\bj k\bl m \bn}
      - {1\over 720} R_{i\bq k \bn} \delta^{p\bq} R_{p\bj m \bl}
      + 2 G_{k i \bl m \bj\bn} + \cc \right) \cr
    &\qquad\qquad\qquad
      (x-z)^k (\bx-\bz)^\bl (x-z)^m (\bx-\bz)^\bn
    \cr
    &
    + \left( {3\over 4} U_{i\bj k\bl m \bn}
      - {1\over 48} R_{i\bq k \bn} \delta^{p\bq} R_{p\bj m \bl} \right)
      (x-z)^k (\bx-\bz)^\bl (x+z)^m (\bx+\bz)^\bn
    \cr
    &
    + \left( {3\over 16} U_{i\bj k\bl m \bn}
      + 2 F_{k i \bl m\bj\bn} \right) 
          (x-z)^k (\bx+\bz)^\bl (x-z)^m (\bx+\bz)^\bn
    + \cc \cr
    &
    + {9\over 16} U_{i\bj k\bl m \bn}
      (x+z)^k (\bx+\bz)^\bl (x+z)^m (\bx+\bz)^\bn
    \cr
    & + \cdots
}}

\subsec{Ricci flatness}

We are now in a position to check the consistency condition \detcond\
at higher order.

On the one hand, we list the constraints on the coefficients $STU$
following from Ricci flatness: given that
\eqn\Ricci{\eqalign{
    0 & \equiv g^{i\bj}(z) R_{i\bj k \bl}(z) \cr
    & =
    \left\{
    \delta^{i\bj} + \delta^{i\bq}\delta^{\bj p}
    R_{p\bq k\bl} z^k \bz^\bl + \cdots
    \right\} \times \cr
    & \qquad
    \bigl\{
    R_{i\bj k \bl}
    - 12 (S_{i\bj k\bl m}  z^m  + \cc )
    - 24 (T_{i\bj k\bl m p}  z^m z^p + \cc) \cr
    & \quad\qquad
    - 36 U_{i\bj k\bl m \bn}  z^m \bz^\bn
    +
    \delta^{p\bq} R_{i \bar{q} k \bn} R_{p\bj m \bl} z^m \bz^{\bn}
    + \cdots
    \bigr\} ,
}}
we have
\eqn\RicciFlat{\eqalign{
    \delta^{i\bj} R_{i\bj k \bl} & = 0, \cr
    \delta^{i\bj} S_{i\bj k\bl m} & = 0, \cr
    \delta^{i\bj} T_{i\bj k\bl m p} & = 0, \cr
    \delta^{i\bj} U_{i\bj k\bl m \bn} & =
    {1\over 36} \delta^{i\bj} \delta^{p\bq}
    \left(
    R_{i \bar{q} k \bn} R_{p\bj m \bl} +
    R_{i \bar{q} k \bl} R_{p\bj m \bn}
    \right),
}}
for $\forall k,\bl,m,\bn,p$.

In order to compute $\det \hat g$ to $O(z,\bz)^4$, let us put
$\hat g_{i\bj} =  \delta_{i\bj} + \hat h_{i\bj}$,
and use the following formula
\eqn\detghat{\eqalign{
    \det \hat g & = \det (1 + \hat h) \cr
    & = \exp \; \tr \log (1+\hat h) \cr
    & = \exp \; (\tr \hat{h} - {1\over 2} \tr \hat{h}^2 + \cdots)
}}
(Note that $\hat{h}$ is $O(z^1\bz^1)$, we can safely neglect higher
powers of $\hat{h}$.) Thus, we need to check
\eqn\hhatCond{
    \tr \hat h - {1\over 2} \tr \hat{h}^2 = O(z,\bz)^5.
    }
Plugging $\hat{h}$ from \ghatsix\ into \hhatCond,
we immediately get
    \eqn\ghatfactorizeone{\eqalign{
    O(z^1\bz^1) \Rightarrow & \quad
    \delta^{i\bj} R_{i\bj k \bl}  = 0, \cr
    O(z^2\bz^1) \Rightarrow & \quad
    \delta^{i\bj} S_{i\bj k\bl m}  = 0, \cr
    O(z^3\bz^1) \Rightarrow & \quad
    \delta^{i\bj} T_{i\bj k\bl m p}  = 0. \cr
}}
from lower order terms as a necessary condition for $\det \hat g \equiv 1$.
Clearly, these follow from Ricci flatness \RicciFlat.

This is not the case for $O(z^2\bz^2)$ terms.
More precisely, the coefficients of the following terms cancel
using the last equation of \RicciFlat:
\eqn\cancelled{\eqalign{
    &(x + z)^k (\bx + \bz)^\bl (x + z)^m (\bx + \bz)^\bn,\cr
    &(x - z)^k (\bx - \bz)^\bl (x + z)^m (\bx + \bz)^\bn.
}}

Those associated with
\eqn\notcancelled{\eqalign{
    &(x - z)^k (\bx - \bz)^\bl (x - z)^m (\bx - \bz)^\bn,\cr
    &(x - z)^k (\bx + \bz)^\bl (x - z)^m (\bx + \bz)^\bn,\cr
    &(x + z)^k (\bx - \bz)^\bl (x + z)^m (\bx - \bz)^\bn
}}
do not automatically cancel, but can be made to
cancel by appropriate choice of the
terms $F$ and $G$:
\eqn\FGexplicit{\eqalign{
    F_{k i \bl m[\bj\bn]} = &
    + {1\over 192} \left(
      R_{i\bq m\bn} \delta^{\bq p} R_{p\bl k \bj} -
      R_{i\bq m\bj} \delta^{\bq p} R_{p\bl k \bn} 
      \right), \cr
    G_{k i \bl m[\bj\bn]} = &
    - {1\over 2880} \left(
      R_{i\bq m\bn} \delta^{\bq p} R_{p\bl k \bj} -
      R_{i\bq m\bj} \delta^{\bq p} R_{p\bl k \bn}
      \right). \cr
}}

In conclusion, one can find a sixth order $K(Z,\bZ)$
such that both the gauge boson mass condition and $\det \hat g \equiv 1$
are satisfied. 

\newsec{Conclusions}

In this work we found actions for $N$ D$p$-branes sitting at points in
a three complex dimensional K\"ahler manifold, satisfying natural
conditions from D-brane physics, notably the enhancement of gauge
symmetry when the D-branes coincide, and the proportionality of the
mass of a string stretched between two D-branes to the shortest geodesic
distance between them.

These actions would be expected to arise as the low energy limit of
D-brane actions derived from string theory for manifolds with weak curvature.
They are also natural starting points for the definition of Matrix theory
\refs{\bfss,\banks}\ on 
target space $\BR^{5-p} \times T^p \times \CM$, where $\CM$ is topologically
trivial but curved, or a subregion of a larger compact manifold.
The condition on the string masses guarantees that the one-loop
quantum effective action will contain a term $v^4/d^{7-p}$, the
leading short distance behavior of the supergravity interaction in this case
\dos.
Now this is not to say that the action as we have computed it so far
is a complete and consistent definition of Matrix theory in this background --
it seems likely that additional terms higher order in commutators as well as in
derivatives would be required to get the physics right -- but rather that
the consistency conditions which we can check at this order can be
satisfied and with a pleasing degree of uniqueness.

We found that the mass condition cannot be satisfied unless the
manifold is Ricci flat, or (combining with the K\"ahler condition)
a Calabi-Yau manifold.  Now at first this might not seem surprising --
certainly we need to start with a consistent closed string theory background to
define sensible open string actions.
However, the approach taken here only used rather general consistency
conditions, and no details of string theory.  The standard string
theory argument, world-sheet conformal invariance expressed as
the RG fixed point condition on the world-sheet sigma model, does not have
any obvious connection with the starting point or the analysis.

Furthermore, the claim that these actions are appropriate starting points
for Matrix theory certainly suggests that we should look for
an argument independent of string theory.  
What we are saying in this
context is that two a priori independent consistency conditions on the physics
--
that the background satisfy the equations of motion, and that the one-loop
quantum corrections reproduce supergravity interactions --
are in fact related.

Another known argument for Ricci flatness of target spaces for brane theories
which may have a closer relation to the present
story is the requirement for kappa symmetry of the covariant supermembrane
action that the background satisfy the supergravity equation of motion
\refs{\BST,\DHIS}.
Now according to the rules of Matrix theory, we can find membrane solutions
of the action, leading to a possible relation; on the other hand we are
necessarily working in light-cone gauge, where kappa symmetry has already
been fixed.

A related argument which could work after gauge fixing is due to
Aharony, Kachru and Silverstein (unpublished).  String theory or M theory
compactified on a Calabi-Yau target will have $N=2$, $d=4$ (or the equivalent
$N=1$, $d=5$) supersymmetry, and the branes will break half of this.
On the other hand, if the space is not Ricci flat, one would argue that
since there is no covariantly constant spinor, there is no unbroken
supersymmetry, and the D-brane theory cannot be supersymmetric.
Thus, given that we assumed that the D-brane theory is supersymmetric,
we should find that the target space has a covariantly constant spinor
and is thus Calabi-Yau.

Although this is an attractive argument, the problem with it (recognized
by AKS as well) is that the true condition for supersymmetry in string
theory is that the target space have zero integrated Ricci curvature (zero
first Chern class).  We expect $\ap$ corrections to the target space metric,
and these can be compatible with supersymmetry, if we also have $\ap$
corrections to the supersymmetry transformation laws, which modify the
condition for an unbroken supersymmetry away from Ricci flatness.

Our result adds to this the statement that the exact proportionality
of the masses of stretched strings to the geodesic distance must also
gain $\ap$ corrections in this case; we know no argument for or against
this in general.  
For example, the $\BC^3/\Gamma$
orbifold models in \dg, which generically describe D-brane propagation
on non-Ricci flat metrics, will not satisfy the isotropic mass condition
in these cases.  On the other hand, if one can tune the ``seed'' metric in
that construction to produce a Ricci flat physical metric, it will be
interesting to go on and implement the mass condition.

It will be interesting to recast our discussion in more geometrical
language, and find axioms which further constrain the action and
determine the higher commutator terms.
The way in which the Ricci-flatness condition arose in our considerations,
through the equation \detcond, suggests that the full
configuration space $\CX_N$ must be Calabi-Yau.

A very interesting question is whether the mass condition is
stable under quantum corrections.  
Since these are non-renormalizable
sigma models, the question seems best defined for $p\le 1$.
For $p=1$, it would appear to be true at one loop, if the full configuration
space is indeed Calabi-Yau.  On the other hand, one might worry that the
known four-loop beta function would violate it.  Perhaps the coupling to
the gauge fields changes this?
For $p=0$, we do not have renormalization in the conventional sense,
but it seems quite possible that some problems involving large $N$ numbers
of D-branes can be treated by a large $N$ renormalization group.
Some comments on this are made in \dstrings; for both types of renormalization,
it will be very interesting to look at quantum corrections in these models.

\medskip
We thank B. de Wit, D.-E. Diaconescu and M. Kontsevich for useful
conversations. 
M.R.D. thanks Cern for hospitality.
A.K. would like to thank Theory Group of
Lawrence Berkeley National Laboratory for hospitality.
H.O. thanks Physics Departments of Rutgers, Harvard Universities,
Laboratoire de Physique Th\'eorique et Hautes Energies
(L.P.T.H.E.) at Universit\'es Pierre et Marie Curie (Paris VI),
and Aspen Center for Physics for hospitality.

M.R.D. is supported in part by DOE grant DE-FG05-90ER40559.
H.O. is supported in part by
NSF grant PHY-95-14797 and DOE grant DE-AC03-76SF00098.
A.K is supported by the fellowship
from the Japanese Ministry of Education, Science and Culture.

\listrefs
\end